\documentclass[10pt,conference,letterpaper]{IEEEtran}

\usepackage{graphicx}
\usepackage{amssymb}
\usepackage{amsmath}
\usepackage{algorithm}
\usepackage{algorithmic}
\usepackage{balance}

\newcommand{\mynotex}[1]{}

\IEEEoverridecommandlockouts

\usepackage{tikz}
\usepackage{textcomp}
\usepackage{hyperref}
\usepackage{lipsum}

\newcommand\copyrighttext{%
  \footnotesize \textcopyright 2019 IEEE. Personal use of this material is permitted. Permission from IEEE must be obtained for all other uses, in any current or future media, including reprinting/republishing this material for advertising or promotional purposes, creating new collective works, for resale or redistribution to servers or lists, or reuse of any copyrighted component of this work in other works. Accepted in IEEE 5th World Forum on Internet of Things (WF-IoT), 15-18 April 2019. 
}
\newcommand\copyrightnotice{%
\begin{tikzpicture}[remember picture,overlay]
\node[anchor=north,yshift=0pt] at (current page.north) {\fbox{\parbox{\dimexpr\textwidth-\fboxsep-\fboxrule\relax}{\copyrighttext}}};
\end{tikzpicture}%
}

\begin{document}

\title{
OAuth 2.0 meets Blockchain for Authorization  in Constrained IoT Environments
\vspace{-0.0in}
}

\author{
Vasilios A. Siris, Dimitrios Dimopoulos, Nikos Fotiou, Spyros Voulgaris, George C. Polyzos \vspace{0.02in}\\
%
Mobile Multimedia Laboratory, Department of Informatics \\
School of Information Sciences \& Technology \\
Athens University of Economics and Business, Greece\\
\{vsiris, dimopoulosd, fotiou, voulgaris, polyzos\}@aueb.gr \vspace{-0.03in}
}

\maketitle
\copyrightnotice

\normalfont

\begin{abstract}
We present models for utilizing blockchain and smart contract technology with the widely used OAuth 2.0  open authorization framework  to provide delegated authorization for constrained IoT devices. The models involve different tradeoffs in terms of privacy, delay, and cost, while  exploiting key advantages of blockchains and smart contracts. These  include linking payments to authorization grants, immutably recording authorization information and policies in smart contracts, and offering resilience through the execution of smart contract code on all blockchain nodes.
\end{abstract}


\section{Introduction}
The goal of the paper is to propose and discuss models for combining the OAuth 2.0  open authorization framework with blockchain and smart contract technology to provide delegated authorization for constrained IoT devices, which have intermittent or no connectivity to the Internet.
%
The motivation for considering the OAuth 2.0 delegated authorization framework is that it is a widely used  IETF standard that is currently being investigated for authorization in IoT environments by IETF's  Authentication and Authorization for Constrained Environments (ACE) Working Group \cite{Sei++16,Sei++19}.
An important feature of OAuth 2.0 is that it provides authorization for different levels of access, termed \emph{scopes}.
Nevertheless, we note that OAuth 2.0 mainly defines the format of the authorization message exchange and the models presented in this paper are applicable for exploiting blockchain and smart contract technology in the general context of authorization in constrained IoT environments.


The remainder of the paper is structured as follows: In Section~\ref{sec:oauth2} we present some background on authorization in constrained environments using the OAuth 2.0 framework. In Section~\ref{sec:blockchains} we present two models for utilizing blockchains and smart contract technology with OAuth 2.0 that involve a different level of integration, and in Section~\ref{sec:evaluation} we present an evaluation of the two models.
Finally, in Section~\ref{sec:related} we present related work and in Section~\ref{sec:conclusions} we present ongoing research extending the models presented in the paper.



\section{Authorization in constrained environments}
\label{sec:oauth2}
\mynotex{Features to highlight:
\begin{itemize}
\item OAuth 2.0 delegates the task of providing authorization to a protected resource,  to an external entity called authorization server.
\item OAuth 2.0 entities and grant types.
\item The format of the access token is transparent to OAuth 2.0.
\item Application of OAuth 2.0 to constrained environments requires modifications that include more efficient encodings of access tokens and PoP (Proof-of-Possession) keys.
\end{itemize}
}
OAuth 2.0 is a framework for delegated authorization to access a protected resource \cite{Har++12}. It enables a third party application (client) to obtain access with specific permissions to a protected resource, with the consent of the resource owner. Access to the resource is achieved through \emph{access tokens},
created by an authorization server.
The specific format of the access tokens, which are discussed in more detail below, is opaque to the clients and to OAuth 2.0.
The authorization consent by the resource owner is provided after the owner is authenticated; however, the authentication procedure is not part of OAuth 2.0. Authorization is provided for different levels of access, such as read and write/modify, which are termed \emph{scopes}, and for a specific time interval.
The OAuth 2.0 authorization flows can involve intermediate messages exchanged before the  access token is provided by the authorization server. The details of the authorization flow does not impact the general approach of the proposed models, hence in our discussion we only consider the initial client request and the authorization server's response containing the access token.

One type of access tokens are \emph{bearer tokens}. Bearer tokens allow the holder (bearer) of the token, independently of its identity, to access the protected resource.
OAuth 2.0 assumes secure communication between the different entities.
 Moreover, it assumes that the protected resource is always connected to the Internet, hence can  communicate with the authorization server to check the validity and scope of the access tokens presented by clients requesting resource access.
Both of the above two requirements are not always possible in constrained environments \cite{Sei++16}.


JSON Web Token (JWT) is an open standard that defines a  compact format to transmit claims  between parties as a JSON object \cite{Jon++15}.
JWTs can use the JSON Web Signature (JWS) structure to allow claims to be digitally signed or integrity protected with a Message Authentication Code (MAC) \cite{Jon++15b}.
Hence, unlike simple bearer tokens, JWT/JWS tokens are self-contained, i.e., they include  all the necessary information for the protected resource to verify their integrity without communicating with the authorization server. Of course, this requires that during its initialization or  configuration phase the protected resource is cryptographically bound with the authorization server.

In constrained environments, in addition to intermittent or no connectivity, the communication between the client and the protected resource is not secure, hence  transmitting bearer tokens or  self-contained JWTs over such insecure links can allow other parties to obtain them through eavesdropping.
For this reason, in constrained environments Proof-of-Possession (PoP) tokens are used \cite{Sei++19}. PoP tokens include a normal access token, such as a JWT/JWS, and a PoP key~\cite{Jon++16}: access to the protected resource is not possible solely with the access token; the PoP key is necessary. Hence, the PoP key must be kept secret and not transmitted in cleartext over insecure links.
Finally, more efficient encoding of JWTs based on CBOR (Concise Binary Object Representation) has been recently  proposed to reduce the amount of data transferred \cite{Sei++19}.

\section{Combining OAuth 2.0 with blockchains}
\label{sec:blockchains}
The advantages from combining OAuth 2.0 with blockchains and smart contracts are the following:
\begin{itemize}
\item OAuth 2.0 typically requires  the resource owner to be online. Combining OAuth 2.0 with blockchains allows authorizations to be linked to payments on the blockchain,  without requiring the online interaction with the resource owner to provide consent.
\item Blockchains can immutably record hashes of the information exchanged during the OAuth 2.0 message flow and cryptographically link authorization grants to payments, providing indisputable receipts  in the case of conflicts.
\item Smart contracts can encode policies in an immutable and  transparent manner. Policies can depend on payments as well as  other IoT events recorded on the  blockchain.
\item Smart contracts run on a distributed platform and typically involve an invocation cost, hence handling  access requests by smart contracts  can protect against DoS attacks that involve a very high  request rate.
\end{itemize}
Asynchronous authorization, where authorization requests are accepted by the authorization server according to policies defined by the resource owner, without requiring synchronous interaction with the resource owner, is supported with ACE-OAuth \cite{Sei++19} and User-Managed Access (UMA)  \cite{Mal++17}.

We present two models which involve a different level of integration with blockchains and smart contracts, and have different tradeoffs in terms of  privacy, delay, and cost:
\begin{itemize}
\item Linking authorization grants to blockchain payments and recording authorization  information on the blockchain.
\item Smart contract for handling authorization requests and encoding authorization policies.
\end{itemize}
%
%
A hash-lock is a cryptographic lock that can be unlocked by revealing a secret whose hash is equal to the lock's value $h$. Unlocking a hash-lock can be one of the conditions for performing a transaction or for executing a smart contract function.
On a single blockchain, a hash-lock  can be linked to an off-chain capability, e.g.,  message decryption, if the hash-lock secret  is the secret key that can decrypt the message.

Time-locks are locks on a blockchain that automatically unlock after an interval has elapsed. The time interval can be measured in absolute time or can be measured in  the number of blocks mined after a specific block.
One usage of time-locks are refunds:  a user (payer) can transfer an amount of currency to a smart contract address, in the form of a deposit. The smart contract can have a function for a second user to transfer the deposit to another account (the payee's account).
However, if the second user never calls this function, then the first user's deposit would be locked indefinitely in the smart contract's account. To avoid such indefinite locking of funds, the smart contract can also include a refund function that allows the first user to transfer the amount deposited from the smart contract account back to the user's account; however, this function can be called only after some time interval, which is the interval in which the second user must transfer the deposit from the smart contract account to the payees account.
Both hash-locks and time-locks are used in the models presented below.

Contracts that include both hash and time-locks are referred to as hash time-locked contracts (HTLCs)~\cite{HTLC}.
Hash and time-locks can be implemented in blockchains with simple scripting capabilities, such as the Bitcoin blockchain, without  requiring elaborate programming capabilities of smart contracts.
Hash time-locked contracts have been used for atomic cross-chain trading  (atomic swaps) \cite{atomic,But16} and for off-chain transactions between two parties that do not trust each other \cite{Poo++16}.

A problem that is not addressed in this paper is how to verify that the IoT device is legitimate or to verify that the IoT device and the authorization server share a common secret; these problems are addressed in \cite{Fot++18}.

In the models presented below, we assume that the Thing (IoT device) providing the resource that the client wants to access is constrained. The client sends a resource access request to the URL of the authorization server  (first model) or to the ABI (Application Binary Interface) of the smart contract that is responsible for handling access to the IoT device (second model). The URL or ABI can be obtained by having the client send a query to the IoT device or using some discovery mechanism.
Finally, in both models we assume that  the client, the resource owner, and the authorization server have an account (public/private key pair) on the blockchain.

\subsection{Linking authorization grants to blockchain payments and recording hashes of authorization information}
\label{sec:model1}
\begin{figure}[tb]
\centering
\includegraphics[width=3.5in]{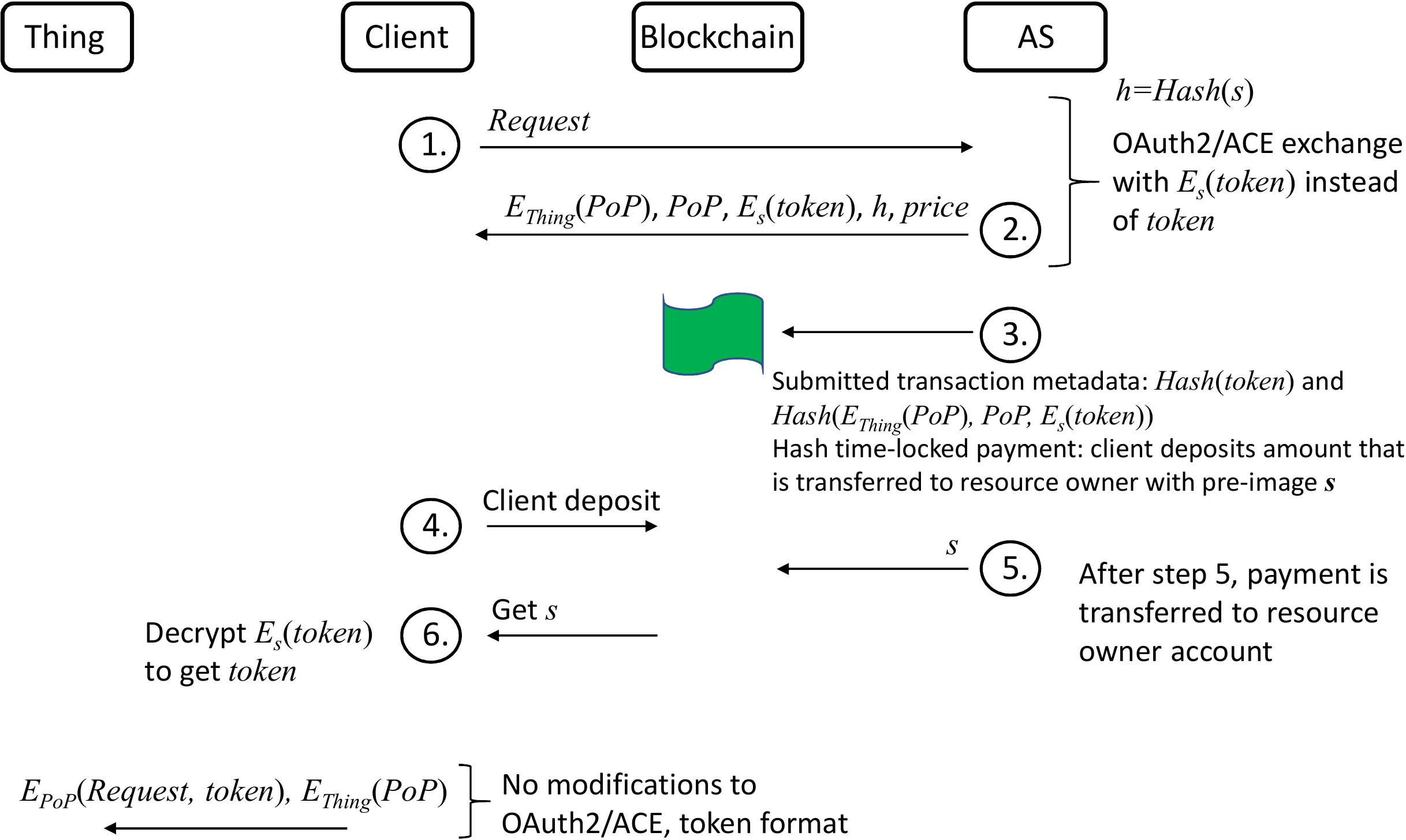}
\caption{Model 1: Authorization grants are linked to blockchain payments and hash of information communicated using OAuth is immutably recorded on the blockchain.}
\label{fig:model1}
\vspace{-0.2in}
\end{figure}

\mynotex{
Key points:
\begin{itemize}
\item Immutable record information exchanged during normal OAuth2 message exchange, that provide verification of authorization grants and non-repudiation in  case of disputes.
\item Either the cleartext of the information exchanges is published or a hash of the information, which represents a tradeoff between transparency and privacy. Level of privacy is also related to whether a public or private/permissioned blockchain is considered.
\item What exactly does recording above information guarantee: 1) client cannot deny receiving specific information contained in OAuth2 message, 2) authorization server cannot deny sending information contained in OAuth2 message.
\item Different access levels can correspond to different prices.
\item Doesn't guarantee that access token client receives provides access to resource.
\end{itemize}
}
With this model the initial communication between the client and the authorization server (AS) occurs as in the normal OAuth 2.0 framework, Figure~\ref{fig:model1}. However, instead of the AS providing the client with authorization credentials after consent is given by the resource owner, the authorization credentials are disclosed after the payment for resource access is recorded on the blockchain. Hence, the resource owner does not need to be  online  to provide consent, as in the case of the normal OAuth 2.0 procedure.

Specifically, in step 1 the client requests resource access from the AS over a secure channel. The AS generates a random  PoP (Proof-of-Possession) key, which in step 2 it sends to the client together with its encryption  with the secret key\footnote{The secret key that the Thing and AS share is established during the configuration phase, when the Thing is bound to the AS.} $Thing$, shared by the Thing and the AS; the client can use this PoP key to establish a secure communication link with the Thing (IoT device). Also, the AS sends to the client the access token encrypted with a secret $s$, $E_s(token)$, the hash $h=Hash(s)$ of the secret $s$, and the price for the level of resource access that is requested. The secret $s$ is a one-time secret randomly generated by the AS for each individual request, and is required for the client to decrypt $E_s(token)$ and obtain the access token; the AS will reveal the secret $s$ once it confirms that the payment for resource access is performed on the blockchain.
The difference with normal OAuth 2.0, in addition to the AS responding immediately to the resource access request without obtaining consent from the resource owner, is that the AS sends the encrypted access token $E_s(token)$ instead of  the access token in plaintext. Also, the AS sends the hash $h$ and the price for resource access. Communicating the price from the AS to the client allows different  levels of resource access, which are encoded in the access token, to correspond to different prices.

In  step 3, two hashes are submitted to the blockchain: the first one is the hash of the  token that the AS will reveal  once payment has been confirmed.
The second one is the  hash of three items: the PoP key encrypted with the secret key the AS shares with the Thing  $E_{Thing}(PoP)$, the PoP key, and the encrypted token $E_s(token)$; the second hash serves as proof of the information that is communicated using OAuth  between the AS and the client.  The two hashes  immutably record on the blockchain the information that has been exchanged, which can  be validated in the case of disputes; however, they do not ensure that the access token  the client obtains from the AS  indeed allows  access to the Thing.

Additionally, in step 3 a hash time-locked payment contract is created on the blockchain, which allows the client to deposit an amount equal to the requested price (step 4). This amount will be transferred to the resource owner's account if the secret $s$ (hash-lock) is submitted to the contract by the AS (step 5) within some time interval. If the time interval is exceeded, then the client can request a refund of the amount it deposited.
Once the secret $s$ is revealed,  the client can obtain $s$ from the blockchain (step 6) and decrypt $E_s(token)$, thus obtain the access token.
At this point, the client has all the necessary information to request access from the Thing, using normal OAuth 2.0 with the modifications from the ACE framework.


\subsection{Smart contracts for handling authorization requests and encoding authorization policies}
\label{sec:model2}
%
In the second model, a smart contract is used to transparently record prices and other authorization policies defined by the resource owner, which is also the owner of the smart contract. Examples of such policies include permitting resource access to specific clients, determined by their public keys on the blockchain, and dependence of access authorization on IoT events that are recorded on the blockchain.
%
\begin{figure}[tb]
\centering
\includegraphics[width=3.5in]{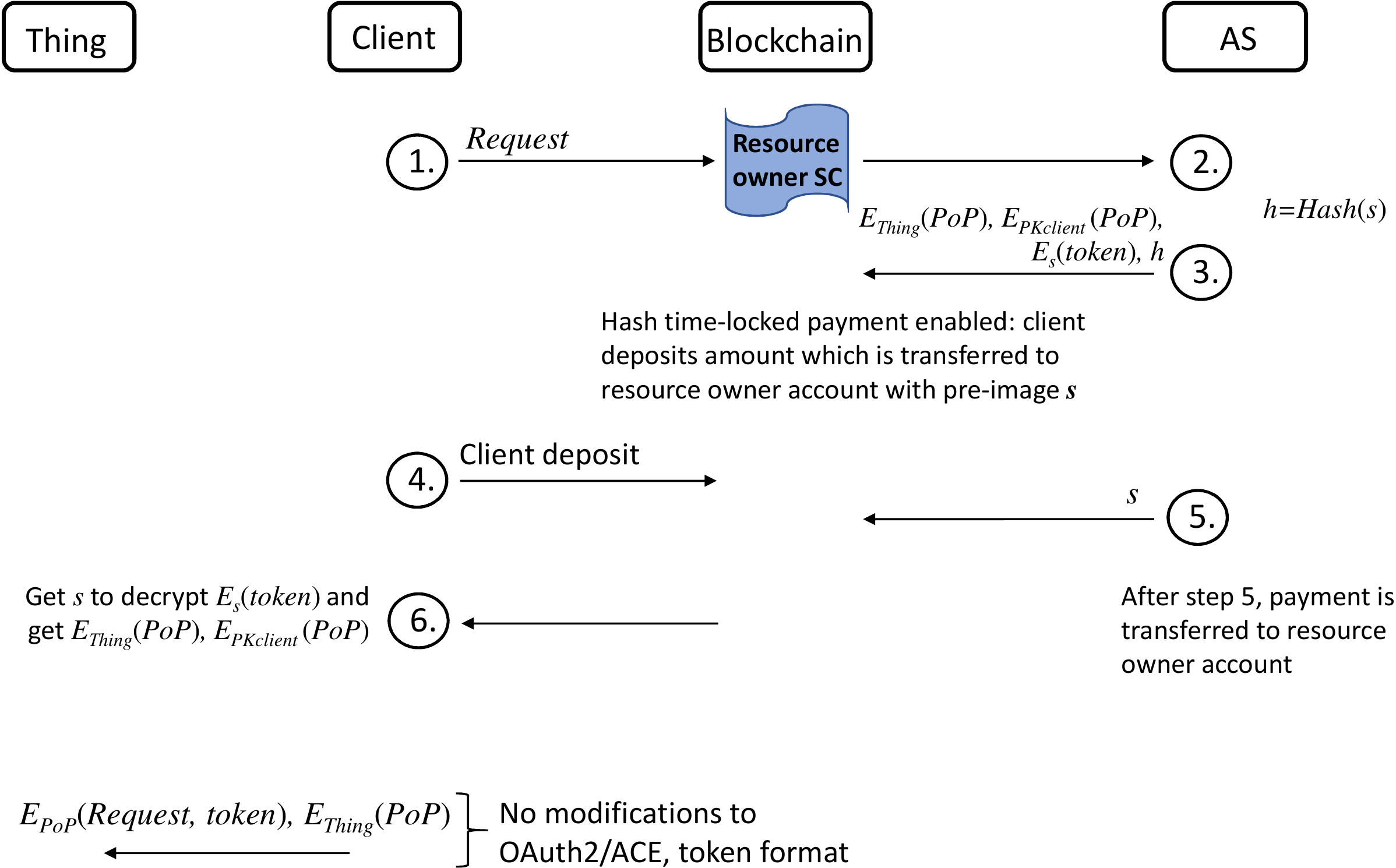}
\caption{Model 2: Smart contract handling authorization requests and encoding authorization policies.}
\label{fig:model2}
\end{figure}

Whereas in the previous model the client and the AS communicated directly, in this model the interaction is through the smart contract, Figure~\ref{fig:model2}.
The smart contract code is executed by all blockchain nodes, providing a secure and reliable execution environment; this provides higher protection against DoS attacks, compared to the model in Section~\ref{sec:model1} where resource access requests are sent directly to the AS.
An additional advantage achieved by allowing a smart contract to handle resource authorization requests is that the smart contract can securely bind the protected resource with the AS responsible for handling authorization requests.

As in the model of Section~\ref{sec:model1}, a hash time-locked payment is enabled, allowing the client to deposit an amount corresponding to the resource access price.
The amount is transferred to the resource owner's account if the secret $s$ that unlocks the hash-lock is revealed.
Once revealed, the client can obtain the secret $s$, together with the other necessary authorization information to access the protected resource.
If the blockchain is public, then $s$ can be read by anyone, hence everyone can obtain the access token. However, the access token cannot be used alone, since the PoP key is also required for accessing the resource. Nevertheless, privacy concerns might require that the token is kept secret; this can be achieved by encoding the token with the client's public key.

In this scenario, the AS sends to the smart contract the PoP key encrypted both with the Thing's key, $E_{Thing}(PoP)$, and with the client's public key, $E_{PKclient}(PoP)$. Hence, only the Thing and the client can obtain the PoP key. On the other hand, in the model of Section~\ref{sec:model1}, the PoP key was sent from the AS to the client over a secure communication link, hence its encryption  was not necessary.


\vspace{-0.01in}
\section{Evaluation}
\label{sec:evaluation}
\vspace{-0.01in}

For the evaluation we have deployed a local  Ethereum node  running  Go-Ethereum\footnote{\texttt{https://geth.ethereum.org/}} that was connected to the Rinkeby\footnote{\texttt{https://www.rinkeby.io/}} public Ethereum testnet.
The local node runs
on a computer with a 4 core CPU at 3.40 GHz, 16 GB RAM, and 64 bit Ubuntu.
Smart contracts were written in Solidity with the Remix\footnote{\texttt{https://remix.ethereum.org/}} web-based editor. The authorization server was based on a PHP implementation of the OAuth 2.0 framework\footnote{\texttt{https://github.com/bshaffer/oauth2-server-php}}. The client used Web3.js to interact with the Rinkeby blockchain.


Table~\ref{tab:results} shows that the second model requires more than three times the amount of gas, hence more than three times the amount of EVM (Ethereum Virtual Machine) resources, compared to the first model; this quantifies the tradeoff between the advantages of the second model, as discussed in Section~\ref{sec:model2}, and its higher cost.
Regarding the delay, Figure~\ref{fig:model1} shows that the first model has three blockchain transactions and Figure~\ref{fig:model2} shows that the second model has four transactions.
Since the total delay is expected to depend mainly on the block mining time, the second model is expected to have a 33\% higher delay for responding to authorization requests.

\begin{table}[tb]
\vspace{-0.05in}
    \caption{Gas  for the two OAuth 2.0 - blockchain models} 
    \centering 
\vspace{-0.1in}
\begin{tabular}{|c|c|} %
        \hline
        &  \footnotesize{Gas}   \\
        \hline 
        \hline
\footnotesize{Model 1: payments and } & 102476  \\
\footnotesize{recording of hashes} &   \\ \hline
\footnotesize{Model 2: smart contract} & 366277  \\
\footnotesize{handling access requests} &   \\ \hline
                 \end{tabular}
    \label{tab:results}
\vspace{-0.2in}
\end{table}

\vspace{-0.01in}
\section{Related work}
\label{sec:related}
\vspace{-0.01in}

The work in \cite{And++17} presents a blockchain-based authorization system where authorization  proofs can be efficiently verified.
The work in \cite{Xu++18} presents a blockchain-based decentralized access control system where IoT devices  interact directly with the blockchain and are always connected, while \cite{Mae++17} presents a system  where policies and access control events are directly recorded on Bitcoin's blockchain.
\cite{Zhang++18} presents a smart contract-based system for providing access control to IoT devices while satisfying access policies in terms of the minimum time interval  between consecutive accesses.
The above works all assume that the IoT device can directly access the blockchain, which is not possible in constrained IoT environments.



The work in \cite{Alp++18b2} presents a system based on  OAuth 2.0 where a smart contract generates authorization tokens, which a key server obtains in order to  provide private keys that allow clients to access a protected resource.
The work of \cite{Har18} contains a high level description for using  smart contracts  with OAuth 2.0 to allow users to freely select the server that  provides authorization to their protected resource.
The difference of this paper  is that we present two different models, with different tradeoffs, for integrating OAuth 2.0 with blockchains,  utilizing hash and time-lock  mechanisms.

\vspace{-0.01in}
\section{Conclusions and future work}
\vspace{-0.01in}
\label{sec:conclusions}

We have presented two models for utilizing blockchain and smart contract technology with the OAuth 2.0 authorization framework, which have different tradeoffs in terms of privacy, delay, and cost.
Ongoing work is investigating using different ledgers for authorizations and payments, and providing decentralized authorization using multiple ASes.

\mynotex{
\begin{itemize}
\item multiple ledgers with hash and time-lock inter-ledger mechanisms
\item decentralized authorization function
\item trusted IoT
\end{itemize}

}

\mynotex{
Extensions and related/possible future work includes the following:
\begin{itemize}
\item The decentralized authorization approach presented in this paper has in some cases similarities with the problem of decentralized oracles, which allow smart contracts to interact with the real world, e.g. by accessing off-chain services through APIs. Specifically, if the interaction involves deterministic queries, such as ``Which city is the capital of Greece?'', then the problem is similar. If however the interaction involves non-deterministic queries, such as ``What is the current temperature in Athens?'', the problem is different.
\item Off-chain transactions to reduce the cost of multiple intermediate payments.
\item Delegated authorization supporting out-of-order authorization,  authorization grants with limited duration; tradeoffs between limited duration authorization grants and revocation; Work on delegated authorization and blockchains is contained in \cite{And++17}.
\end{itemize}
}

\vspace{-0.01in}
\section*{Acknowledgements}
\vspace{-0.01in}
This research  has been undertaken in the context of project
SOFIE (Secure Open Federation for Internet Everywhere), which has received funding from EU's Horizon 2020 programme, under grant agreement No. 779984.


\balance
\bibliographystyle{IEEEtran}

{
\bibliography{IEEEabrv,auth} }

\begin{thebibliography}{10}
\providecommand{\url}[1]{#1}
\csname url@samestyle\endcsname
\providecommand{\newblock}{\relax}
\providecommand{\bibinfo}[2]{#2}
\providecommand{\BIBentrySTDinterwordspacing}{\spaceskip=0pt\relax}
\providecommand{\BIBentryALTinterwordstretchfactor}{4}
\providecommand{\BIBentryALTinterwordspacing}{\spaceskip=\fontdimen2\font plus
\BIBentryALTinterwordstretchfactor\fontdimen3\font minus
  \fontdimen4\font\relax}
\providecommand{\BIBforeignlanguage}[2]{{%
\expandafter\ifx\csname l@#1\endcsname\relax
\typeout{** WARNING: IEEEtran.bst: No hyphenation pattern has been}%
\typeout{** loaded for the language `#1'. Using the pattern for}%
\typeout{** the default language instead.}%
\else
\language=\csname l@#1\endcsname
\fi
#2}}
\providecommand{\BIBdecl}{\relax}
\BIBdecl

\bibitem{Sei++16}
L.~Seitz \emph{et~al.}, ``{Use Cases for Authentication and Authorization in
  Constrained Environments},'' RFC 7744, IETF, January 2016.

\bibitem{Sei++19}
------, ``{Authentication and Authorization for Constrained Environments (ACE)
  using the OAuth 2.0 Framework (ACE-OAuth)},'' IETF Draft, February 14, 2019.

\bibitem{Har++12}
D.~Hardt \emph{et~al.}, ``{The OAuth 2.0 Authorization Framework},'' RFC 6749,
  Standards Track, IETF, October 2012.

\bibitem{Jon++15}
M.~Jones, J.~Bradley, and N.~Sakimura, ``{JSON Web Token (JWT)},'' RFC 7519,
  Standards Track, IETF, May 2015.

\bibitem{Jon++15b}
------, ``{JSON Web Signature (JWS)},'' RFC 7515, Standards Track, IETF, May
  2015.

\bibitem{Jon++16}
M.~Jones, J.~Bradley, and H.~Tschofenig, ``{Proof-of-Possession Key Semantics
  for JSON Web Tokens (JWTs)},'' RFC 7800, Standards Track, IETF, April 2016.

\bibitem{Mal++17}
\BIBentryALTinterwordspacing
E.~Maler \emph{et~al.}, ``{User-Managed Access (UMA) 2.0 Grant for OAuth 2.0
  Authorization},'' May 25, 2017, last accessed 24/02/2019. [Online].
  Available: \url{https://docs.kantarainitiative.org/uma/wg/oauth-uma-grant-2.0-05.html}
\BIBentrySTDinterwordspacing

\bibitem{HTLC}
{Bitcoin Wiki}, ``{Hashed Timelock Contracts (HTLC)},''
  https://en.bitcoinwiki.org/wiki/Hashed\_Timelock\_Contracts, last accessed
  24/02/2019.

\bibitem{atomic}
------, ``{Atomic cross-chain trading},''
  https://en.bitcoinwiki.org/wiki/Atomic\_cross-chain\_trading, last accessed
  24/02/2019.

\bibitem{But16}
V.~Buterin, ``{Chain Interoperability},'' R3 Report, September 2016.

\bibitem{Poo++16}
J.~Poon and T.~Dryja, ``{The Bitcoin Lightning Network: Scalable oﬀ-chain
  instant payments},'' https://lightning.network/lightning-network-paper.pdf,
  January 14, 2016, last accessed 24/02/2019.

\bibitem{Fot++18}
N.~Fotiou, V.~A. Siris, and G.~C. Polyzos, ``{Interacting with the Internet of
  Things using Smart Contracts and Blockchain Technologies},'' in \emph{Proc.
  of 7th Int'l Symp. on Security \& Privacy on Internet of Things, in
  conjunction with SpaCCS}, 2018.

\bibitem{And++17}
M.~P. Andersen \emph{et~al.}, ``{WAVE: A Decentralized Authorization System for
  IoT via Blockchain Smart Contracts},'' University of California at Berkeley,
  Tech. Rep., December 2017.

\bibitem{Xu++18}
R.~Xu \emph{et~al.}, ``{BlendCAC: A BLockchain-ENabled Decentralized
  Capability-based Access Control for IoTs},'' arXiv:1804.09267v1, April 2018.

\bibitem{Mae++17}
D.~D.~F. Maesa, P.~Mori, and L.~Ricci, ``Blockchain based access control,'' in
  \emph{Proc. of 17th {IFIP} Distributed Applications and Interoperable Systems
  (DAIS)}, 2017.

\bibitem{Zhang++18}
Y.~Zhang \emph{et~al.}, ``{Smart Contract-Based Access Control for the Internet
  of Things},'' arXiv:1802.04410, February 2018.

\bibitem{Alp++18b2}
O.~Alphand \emph{et~al.}, ``{IoTChain: {A} blockchain security architecture for
  the Internet of Things},'' in \emph{Proc. of {IEEE} WCNC}, 2018.

\bibitem{Har18}
\BIBentryALTinterwordspacing
T.~Hardjono, ``{Decentralized Service Architecture for OAuth2.0},'' March 25,
  2018. [Online]. Available: \url{https://tools.ietf.org/html/draft-hardjono-oauth-decentralized-02}
\BIBentrySTDinterwordspacing

\end{thebibliography}

\end{document}